\documentclass[10pt, conference, compsocconf]{IEEEtran}
\usepackage{mathptmx} 

\usepackage{graphicx}
\usepackage{fancyhdr}
\usepackage[normalem]{ulem}
\usepackage[hyphens]{url}
\usepackage[sort,nocompress]{cite}
\usepackage[final]{microtype}
\usepackage[keeplastbox]{flushend}
\usepackage[bookmarks=true,breaklinks=true,letterpaper=true,colorlinks,citecolor=blue,linkcolor=blue,urlcolor=blue]{hyperref}

\usepackage{color,xcolor}
\usepackage{xspace}
\usepackage{multirow}
\usepackage{booktabs}
\usepackage{lipsum}
\usepackage{soulutf8}
\usepackage{diagbox}
\usepackage{physics}

\usepackage{colortbl}

\usepackage{enumitem}
\usepackage{braket}
\usepackage{calc}

\usepackage{amsmath}
\usepackage{amsfonts}
\usepackage{url}

\usepackage{afterpage}
\usepackage{cancel}

\usepackage[ruled,vlined]{algorithm2e}
\usepackage{setspace}

\usepackage{pifont}

\usepackage{titlesec}

\usepackage{hyperref}
\hypersetup{
    colorlinks=true,
    linkcolor=magenta,
    filecolor=magenta,      
    urlcolor=blue,
}

\usepackage{xcolor}
\usepackage{soul}

\definecolor{citecolor}{RGB}{34,139,34}
\definecolor{mydarkblue}{rgb}{0,0.08,1}
\definecolor{mydarkgreen}{rgb}{0.02,0.6,0.02}
\definecolor{mydarkred}{rgb}{0.8,0.02,0.02}
\definecolor{mydarkorange}{rgb}{0.40,0.2,0.02}
\definecolor{mypurple}{RGB}{111,0,255}
\definecolor{myred}{rgb}{1.0,0.0,0.0}
\definecolor{mygold}{rgb}{0.75,0.6,0.12}
\definecolor{myblue}{rgb}{0,0.2,0.8}
\definecolor{mydarkgray}{rgb}{0.,0.2,0.2}

\definecolor{lightred}{RGB}{255,235,235}
\definecolor{lightgreen}{RGB}{235,255,235}
\definecolor{lightblue}{RGB}{235,235,255}
\definecolor{lightcyan}{RGB}{235,255,255}
\definecolor{lightmagenta}{RGB}{255,235,255}
\definecolor{lightyellow}{RGB}{255,255,235}

\definecolor{qxkcolor}{RGB}{215,235,255}
\definecolor{softmaxcolor}{RGB}{230,235,255}
\definecolor{probxvcolor}{RGB}{255,255,235}

\definecolor{topkcolor}{RGB}{255,235,235}
\definecolor{zecolor}{RGB}{255,255,235}
\definecolor{dynacolor}{RGB}{235,255,255}

\definecolor{reviewcolor}{RGB}{0,0,200}

\renewcommand\footnotemark{}

\newcommand\blfootnote[1]{%
  \begingroup
  \renewcommand\thefootnote{}\footnote{#1}%
  \addtocounter{footnote}{-1}%
  \endgroup
}

\newcommand{\name}{RobustState\xspace}

\newcommand{\block}{\textit{block}}

\newcommand{\x}{$\times$\xspace}

\newcommand{\cnot}{\texttt{CNOT}}
\newcommand{\crgate}{\texttt{CR}}

\newcommand{\swap}{\texttt{SWAP}}
\newcommand{\rx}{\texttt{RX}\xspace}
\newcommand{\rzx}{\texttt{RZX}\xspace}
\newcommand{\ry}{\texttt{RY}\xspace}
\newcommand{\rz}{\texttt{RZ}\xspace}

\newcommand{\sx}{\texttt{SX}}

\newcommand{\xgate}{\texttt{X}}

\newcommand{\cmark}{\ding{51}}%
\newcommand{\xmark}{\ding{55}}%

\newcounter{rlabelno}

\title{\name: Boosting Fidelity of Quantum State Preparation via Noise-Aware Variational Training} 

\author{\IEEEauthorblockN{Hanrui Wang*$^{1}$, Yilian Liu*$^{1}$, Pengyu Liu*$^{2}$, Jiaqi Gu$^{3}$, Zirui Li$^{1}$, Zhiding Liang$^{4}$, Jinglei Cheng$^{5}$,\\
Yongshan Ding$^{6}$, Xuehai Qian$^{5}$, Yiyu Shi$^{4}$, David Z. Pan$^{3}$, Frederic T. Chong$^{7}$, Song Han$^{1}$}
\IEEEauthorblockA{$^{1}$MIT $^{2}$CMU $^{3}$University of Texas at Austin $^{4}$University of Notre Dame \\ $^{5}$Purdue University $^6$Yale University $^7$University of Chicago}
}

\begin{document}
\maketitle
\pagestyle{plain}

\begin{abstract}

Quantum state preparation, a crucial subroutine in quantum computing, involves generating a target quantum state from initialized qubits. Arbitrary state preparation algorithms can be broadly categorized into arithmetic decomposition (AD) and variational quantum state preparation (VQSP). AD employs a predefined procedure to decompose the target state into a series of gates, whereas VQSP iteratively tunes ansatz parameters to approximate target state. VQSP is particularly apt for Noisy-Intermediate Scale Quantum (NISQ) machines due to its shorter circuits. However, achieving noise-robust parameter optimization still remains challenging.

There exist two types of noise-aware optimizers: gradient-free and gradient-based. Gradient-free optimizers, such as Bayesian optimization (BO), update parameters without gradient information, treating the problem as a black box. Gradient-based optimizers, on the other hand, utilize gradients derived from the parameter shift (PS) rule. Both approaches suffer from \textit{low training efficiency}: gradient-free optimizers lack accurate gradient guidance, and PS demands a costly $\mathcal{O}(N)$ executions for $N$ parameters. While existing noise-unaware optimizers can exploit back-propagation on a classical simulator for efficient gradient computation in a \textit{single} backward pass, they exhibit \textit{low robustness} on real machines.

We present RobustState, a novel VQSP training methodology that combines high robustness with high training efficiency. The core idea involves utilizing measurement outcomes from real machines to perform back-propagation through classical simulators, thus incorporating real quantum noise into gradient calculations. RobustState serves as a versatile, plug-and-play technique applicable for training parameters from scratch or fine-tuning existing parameters to enhance fidelity on target machines.
It is adaptable to various ansatzes at both gate and pulse levels and can even benefit other variational algorithms, such as variational unitary synthesis.

Comprehensive evaluation of RobustState on state preparation tasks for 4 distinct quantum algorithms using 10 real quantum machines demonstrates a coherent error reduction of up to 7.1 $\times$ and state fidelity improvement of up to 96\% and 81\% for 4-Q and 5-Q states, respectively. On average, RobustState improves fidelity by 50\% and 72\% for 4-Q and 5-Q states compared to baseline approaches.

\end{abstract}

\section{Introduction}
\label{sec:intro}

\begin{figure}[t]
    \centering
    \includegraphics[width=\columnwidth]{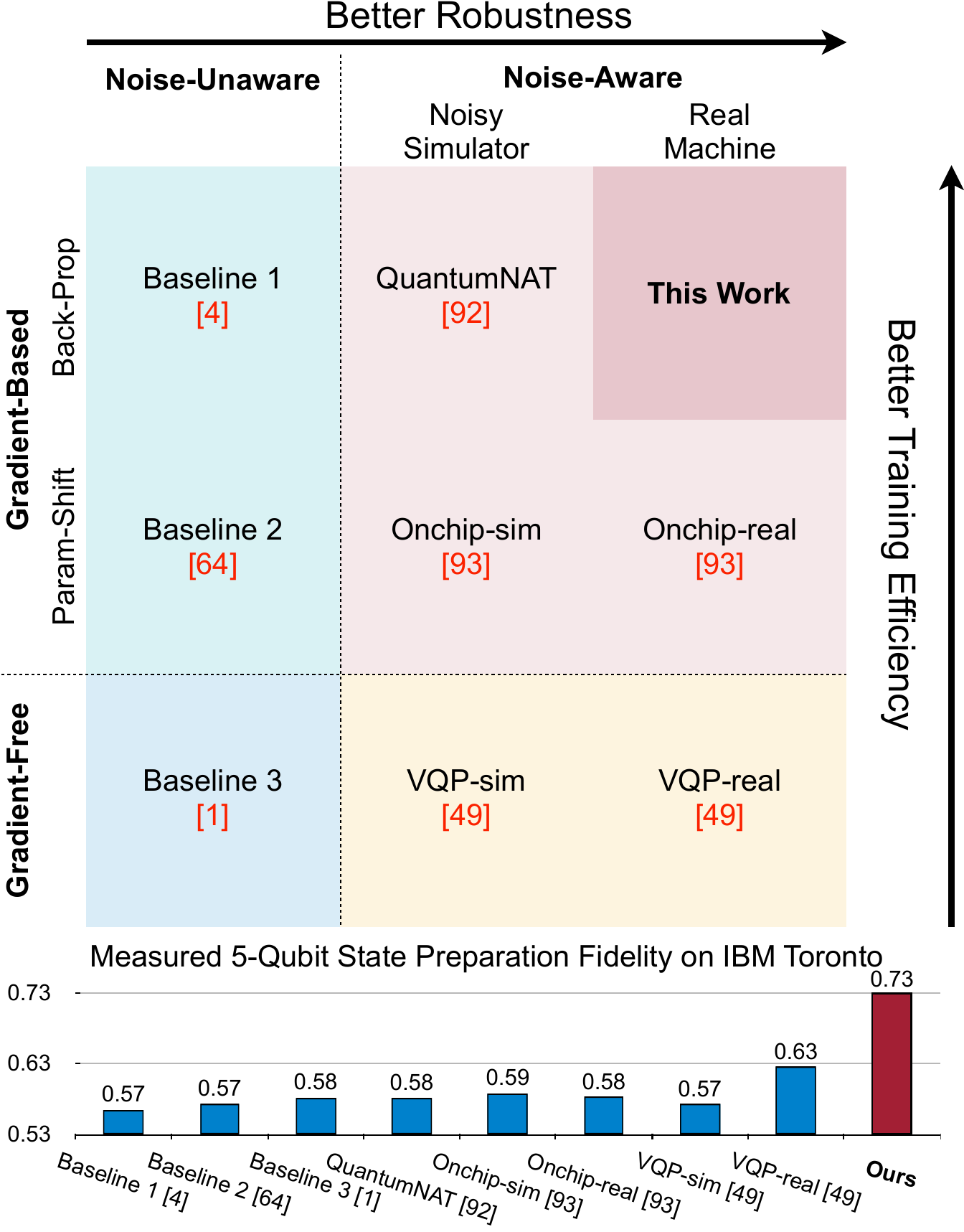}
    \caption{Proposed \name performs back-propagation training using results from real quantum machines, achieving high robustness and training efficiency. State preparation fidelity of each method is evaluated on real machines.}
    \label{fig:teaser}
    \vspace{-5pt}
\end{figure}

\begin{figure}[t]
    \centering
    \includegraphics[width=\columnwidth]{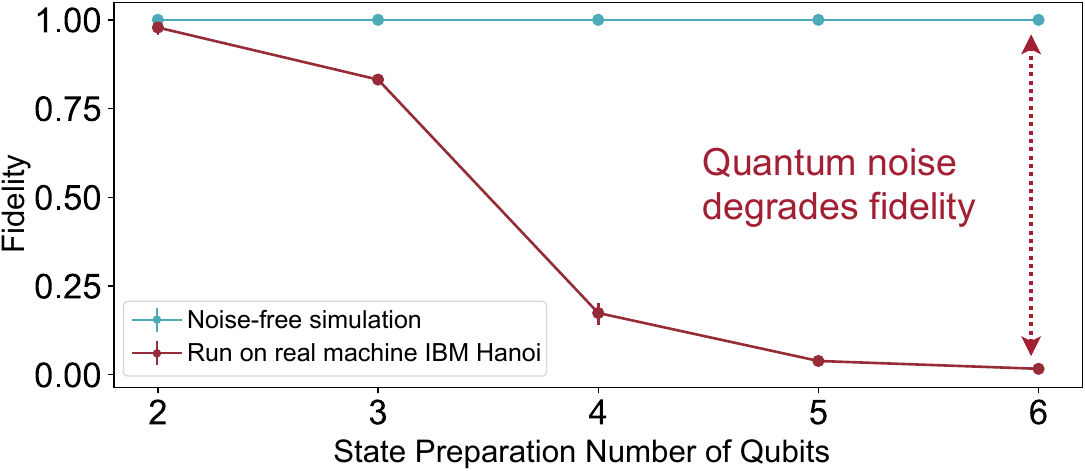}
    \caption{Prepared state fidelity on a noise-free simulator and real machine IBM Hanoi. (i) Noise-free fidelity is very close to 100\%. (ii) Noise significantly reduces fidelity, especially for large qubit numbers.}
    \label{fig:large_noise}
\end{figure}

\blfootnote{*Equal Contributions.} Quantum Computing (QC) is emerging as a promising computing paradigm, garnering significant research interest for addressing previously unsolvable problems with enhanced efficiency. Various industries and disciplines stand to benefit from QC, including cryptography~\cite{shor1999polynomial}, database search~\cite{grover1996fast}, combinatorial optimization~\cite{farhi2014quantum}, molecular dynamics~\cite{peruzzo2014variational}, and machine learning~\cite{biamonte2017quantum, lloyd2013quantum, rebentrost2014quantum, liang2021can}, among others. Progress in physical implementation technologies has spurred rapid advancements in QC hardware over the past two decades, leading to the recent release of multiple QC systems with up to 433 qubits~\cite{ibm127, rigetti, google72, intel49}.

Quantum state preparation, a crucial subroutine in QC, facilitates the preparation of the system's initial state. This process is essential for applications such as codewords in quantum error correction~\cite{horsman2012surface}, amplitude encoding~\cite{nakaji2022approximate} in quantum machine learning, and initial condition loading for solving Partial Differential Equations (PDEs) using quantum machines~\cite{Lubasch_2020, gonzalezconde2022simulating}.

State preparation can be achieved through two primary approaches: arithmetic decomposition (AD) and variational quantum state preparation (VQSP). AD methods, such as Shannon decomposition~\cite{madden2022} and Mottonen decomposition~\cite{mottonen2004}, focus on generating circuits to prepare the target state. Shannon decomposition, inspired by Shannon's expansion theorem in classical reversible computing, constructs a quantum circuit by first synthesizing a series of classical reversible gates and subsequently translating them into quantum gates~\cite{madden2022}. Mottonen decomposition, in contrast, decomposes the target state into single-qubit rotations and multi-qubit controlled rotations~\cite{mottonen2004}. This process iteratively constructs the target state hierarchically, beginning with the least significant qubit and progressing with each layer of controlled rotations acting on an increasing number of qubits.
The second approach, variational quantum state preparation (VQSP)~\cite{arrazola2019machine, cincio2021machine}, employs a different strategy by iteratively updating parameters in a variational circuit ansatz. This iterative process aims to minimize the distance between the implemented final state and the target state.

On NISQ machines, AD methods can be significantly affected by quantum noise. Utilizing IBM Qiskit\cite{qiskit} compiler for Shannon decomposition, we generate circuits that transform the all-zero state into three distinct quantum states and compare the average fidelity on noise-free simulators and real quantum machines (Fig.~\ref{fig:large_noise}). Although the compiler can achieve 100\% fidelity in noise-free simulations, real quantum machines experience considerable fidelity degradation. This fidelity gap intensifies with an increasing number of qubits, e.g., while 2-qubit (2-Q) states achieve 98\% fidelity on real machines, 6-qubit (6-Q) states have a mere 1\% fidelity. Conversely, VQSP is better suited for NISQ devices due to its flexibility in mitigating coherent errors by adjusting circuit parameters and its reduced 2-qubit gate count and depth, as observed in our experiments. 

Nevertheless, Optimizing VQSP parameters for \textit{noise robustness} remains a formidable challenge. Fig.\ref{fig:teaser} presents a taxonomy of existing optimization strategies for parameter training and their performance on a 5-qubit (5-Q) quantum state on the IBM Toronto machine. Although noise-free fidelities exceed 99\%, fidelity on real machines significantly declines. The performance of noise-unaware simulation methods in the first column, such as training on a noise-free simulator, deteriorates due to real machine noise. In the second column, training with noisy simulators like QuantumNAT\cite{wang2021quantumnat}, onchip-sim~\cite{qocdac2022}, and VQP-sim~\cite{liang2022variational}, still results in low state fidelity on real devices due to discrepancies between the simulator and the actual device. Despite that, training on classical differentiable noise-free or noisy simulators is \textit{faster than black-box} optimizers for achieving the same fidelity level, as accurate gradients for all parameters can be computed in a \textit{single} backward pass. The rightmost column outlines optimization methods using real machines. Gradient-free optimizers, such as Bayesian optimization, update parameters with reward feedback from real devices, while gradient-based optimizers update parameters with gradients computed from the parameter shift (PS) rule executed on real machines. Regrettably, both methods suffer from low training efficiency. Gradient-free optimizers treat the problem as a black box, lacking accurate gradient guidance, and PS, despite providing gradient guidance, incurs high costs, necessitating $\mathcal{O}(\#\text{params})$ executions for a single parameter update round.

The natural goal of achieving both high robustness and training efficiency calls for \textit{noise-aware back-propagation}. However, due to the No-Cloning Theorem~\cite{dieks1982communication, wootters1982single}, back-propagation on real quantum machines is infeasible, as intermediate results (quantum states) cannot be stored for use in the backward pass.

To address this challenge, we introduce \name. The core idea hinges on the fact that while real quantum machines cannot provide intermediate results, they can supply \textit{final results} through measurements. Thus, we can use final results from real quantum machines combined with intermediate results from classical simulators to complete the backward pass. In a single training step, the same set of parameters is executed on \textit{both} real quantum machines and simulators. The loss function is computed between the tomography states from a real machine and the target state. Subsequently, noise-impacted gradients are back-propagated through the simulator using the previously simulated intermediate results, ensuring noise resilience in the trained parameters. Fig.~\ref{fig:teaser} (bottom) highlights the superior accuracy of our method over alternatives. The two primary advantages of our approach are summarized in Table~\ref{tab:optmethod}.

RobustState balances increased classical computation in simulators against reduced executions on real devices and enhanced fidelity by obtaining gradients for all parameters with one tomography. While classical simulation scalability may be a concern, we argue the following points: (i) The scalability of our approach is \textit{comparable} to that of state-of-the-art arithmetic decomposition methods~\cite{madden2022, mottonen2004}, yet it attains significantly higher fidelity. (ii) Preparing small to medium-sized states with high fidelity is a crucial task in quantum computing. For instance, the five-qubit code~\cite{gottesman2010introduction} and nine-qubit surface code~\cite{fowler2012surface} are essential starting procedures for all programs in fault-tolerant quantum computing~\cite{ravi2022have, 10.1145/3470496.3527381, 10.1145/3470496.3527417, das2022afs, ueno2022qulatis}. (iii) Block-wise optimized quantum circuit compilers, such as~\cite{xu2022quartz}, benefit significantly from generating unitaries for a few qubits with high fidelity, as this serves as a crucial subroutine for improving the final circuit fidelity. Thus the potential scalability concerns of classical simulators are outweighed by the value and importance of preparing small to medium-sized states with high fidelity.

\name offers a plug-and-play approach capable of enhancing the fidelity of any existing ansatz, such as those from Xanadu~\cite{Lubasch_2020} and QuantumNAS~\cite{wang2022quantumnas}. Unlike traditional PS rules, which necessitate specific quantum gate structures~\cite{mitarai2018quantum} for gradient computation, \name calculates gradients through a simulator, eliminating the need for particular ansatz structures and ensuring broad applicability to any classically simulatable ansatz. Consequently, our method can be employed to improve the performance of \textit{both gate and pulse} ansatzes, as demonstrated in our experiments. Moreover, the noise-aware back-propagation technique can be extended to other variational algorithms, highlighting its wide-ranging applicability and effectiveness. Our experiments involving variational unitary synthesis and quantum state regression provide further evidence of its extensive utility. In summary, \name makes the following contributions:

\indent \textbf{$\bullet$ Noise-aware gradient back-propagation approach:} This method enhances both the robustness and training efficiency of variational state preparation circuits.

\indent \textbf{$\bullet$ Extensive applicability across various ansatzes and applicable to other variational algorithms:} \name is applicable at both gate and pulse levels and is compatible with existing ansatz design and search frameworks. The methodology can also be applied to other variational algorithms such as unitary synthesis and quantum state regression.

\indent \textbf{$\bullet$ Comprehensive experiments on 10 real machines:} Our results demonstrate that \name improves 4-Q state fidelity by \textbf{50\%} and 5-Q state fidelity by \textbf{72\%} on average, reducing coherent errors by up to \textbf{7.1$\times$} when compared to noise-unaware baselines.

\begin{table}[t]
\centering
\renewcommand*{\arraystretch}{1}
\setlength{\tabcolsep}{1pt}
\footnotesize

\begin{tabular}{lcccc}
\toprule
 &Parameter-Shift & Gradient-Free & \name \\
\midrule
Scaling w.r.t. \#Params&  $\mathcal{O}(n)$ & Unscalable & $\mathcal{O}(1)$\\
Gradient Guidance & \cmark & \xmark & \cmark\\ 

\bottomrule

\end{tabular}

\caption{Comparison between different noise-aware optimizers.}

\label{tab:optmethod}

\end{table}

\section{Variational Quantum State Preparation}
In this section, we briefly introduce several concepts used in variational quantum state preparation.

\textbf{Quantum state preparation.} Grover and Rudolph initially introduced the quantum state preparation problem for efficiently generating integrable probability distribution functions~\cite{grover2002}. Numerous state preparation techniques have been proposed since, targeting specific states like Gaussian wave functions~\cite{kitaev2008, bauer2021, Rattew2021efficient}, continuous functions~\cite{Rattew2022, holmes2020}, and arbitrary functions~\cite{zhang2022, Sun:2021xof, sanchez2021, plesch2011}. State preparation techniques can be categorized into Arithmetic Decomposition (AD) and Variational Quantum State Preparation (VQSP). AD utilizes rule-based algorithms to generate a circuit mapping the $\ket{0}$ state to the target state $\ket{\psi}$ in one step, while VQSP iteratively refines a circuit to minimize the difference between the produced and target states. VQSP begins with designing a parameterized circuit architecture (called ansatz). The circuit realizes a parameterized unitary $U(x,\theta)$, preparing a state: $\ket{\psi(x, \theta)} = U(x, \theta)\ket{0\dotsc 0}$, where $x$ is the input data for computation, and $\theta$ is a set of free variables for adaptive optimizations. A set of circuit parameters are then trained using a hybrid quantum-classical optimization procedure, iteratively updating parameters in $U(x,\theta)$ to minimize loss.

\textbf{Hardware-efficient ansatz.} Although AD can prepare arbitrary quantum states precisely in theory, practical application is limited on real machines due to topological constraints and significant hardware noise. Exact preparation schemes require all-to-all connectivity between qubits, so additional \swap\xspace gates are needed for real-machine deployment. To address these issues, researchers proposed hardware-efficient ansatzes~\cite{kandala2017hardware, gokhale2020optimized} that use only native (or near-native) gates provided by a specific backend to produce compact, expressive circuits. More compact circuits are more robust against noise, so we base our framework on variational quantum circuits to tackle topological constraints and hardware noise.

\textbf{Quantum state fidelity.}
Quantum state fidelity measures how closely two states match, defined as $F(\rho, \sigma)=(\tr \sqrt{\sqrt{\rho} \sigma \sqrt{\rho}})^{2}$, where $\sqrt{\rho}>0$ and $(\sqrt{\rho})^2=\rho$. If either $\rho$ or $\sigma$ is a pure state, fidelity simplifies to $F(\rho, \sigma)=\tr(\rho\sigma)$. If $F(\rho, \sigma)=1$, $\rho$ and $\sigma$ are identical. We will use fidelity as a prepared state quality indicator throughout this paper.

\textbf{Incoherent and coherent noises.}
On real quantum computers, errors arise from qubit-environment interactions and imprecise controls \cite{krantz2019quantum, bruzewicz2019trapped, magesan2012characterizing}. Qubit-environment interactions lead to quantum state information loss, termed \textit{incoherent errors}, expressed mathematically as $\ketbra{\phi}\to (1-p)\ketbra{\phi}+p\frac{I}{d}$, where $p$ represents incoherent error strength. Imprecise quantum gate controls cause coherent errors, where the state vector deviates from the ideal state vector: $\ketbra{\phi}\to \ketbra{\phi'}$.

Coherent errors can be mitigated through software approaches. Characterization~\cite{magesan2012characterizing} and calibration~\cite{ibm_2021} are two standard methods to estimate and reduce coherent errors. However, even with calibration, coherent errors cannot be entirely eliminated. In state preparation, we model errors as a combination of incoherent and coherent errors. Assuming we aim to prepare state $\rho=\ketbra{\phi}$ but obtain $\rho'=(1-p)\ketbra{\phi'}+p\frac{I}{d}$, where $p$ signifies incoherent error strength, and $\ket{\phi}\neq \ket{\phi'}$ indicates coherent error. We quantify coherent error using the infidelity between $\ket{\phi}$ and $\ket{\phi'}$. With simple derivations, we can approximate coherent error as:
\begin{equation}
1-\frac{\tr(\rho\rho')}{\sqrt{\tr(\rho^2)}}
\end{equation}


\section{Noise-Aware \name Methodology}

In this section, we first discuss the advantages of using variational state preparation over conventional arithmetic decomposition. Next, we introduce the \name approach for noise-aware back-propagation parameter training. Finally, we discuss the extensive applicability of \name on both gate and pulse ansatzes.

\subsection{Arithmetic Decomposition versus Variation State Preparation}

\begin{figure}[t]
    \centering
    \includegraphics[width=\columnwidth]{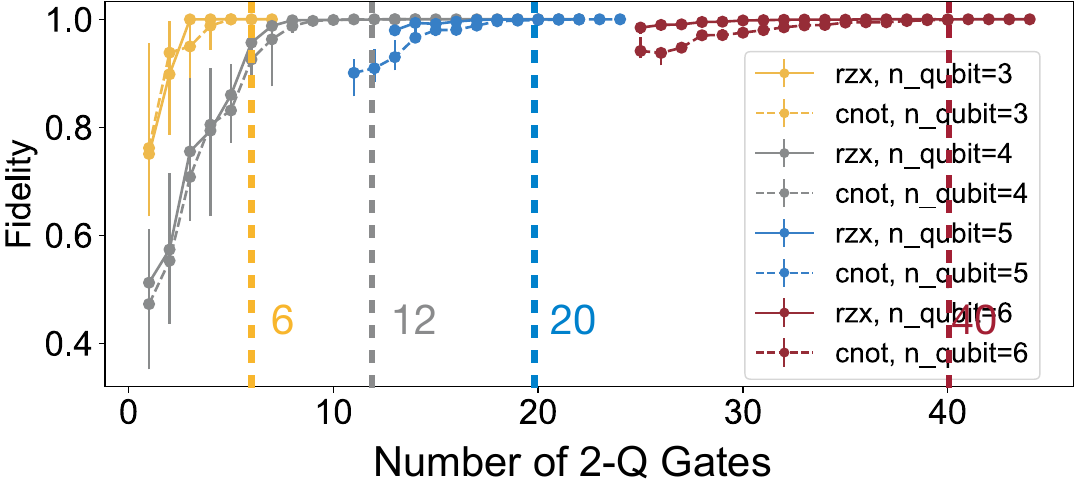}
    \caption{Noise-free state fidelity for 3-Q, 4-Q, 5-Q, and 6-Q states versus the number of 2-Q gates. Each ansatz is tested with both \cnot\xspace and \rzx as the 2-Q gate.}
    \label{fig:n2q_vs_fidelity}
    \vspace{-5pt}
\end{figure}

We conduct experimental comparisons between arithmetic decomposition and variational approaches. For AD, we employ Qiskit's \texttt{initialize} method and the Mottonen decomposition method~\cite{mottonen2004}. For VQSP, we choose ansatzes with sufficient expressibility by examining the fidelity when using varying numbers of 2-Q gates. As the number of 2-Q gates increases, the ansatz's expressibility becomes adequate for generating any quantum state. We use hardware-efficient ansatzes to ensure no extra \swap\xspace gates are required. As demonstrated in Fig.~\ref{fig:n2q_vs_fidelity}, we ensure the ansatz converges to the target state by selecting $6$ 2-Q gates for 3-Q states, $12$ for 4-Q states, and $20$ for 5-Q states. Table~\ref{tab:fid} presents the mean and standard deviation for the fidelity of chosen ansatzes for $100$ randomly generated states. The means all exceed $99$\% with standard deviations smaller than $0.002$, indicating the ansatzes' sufficient expressibility.

\begin{figure}[t]
    \centering
    \includegraphics[width=\columnwidth]{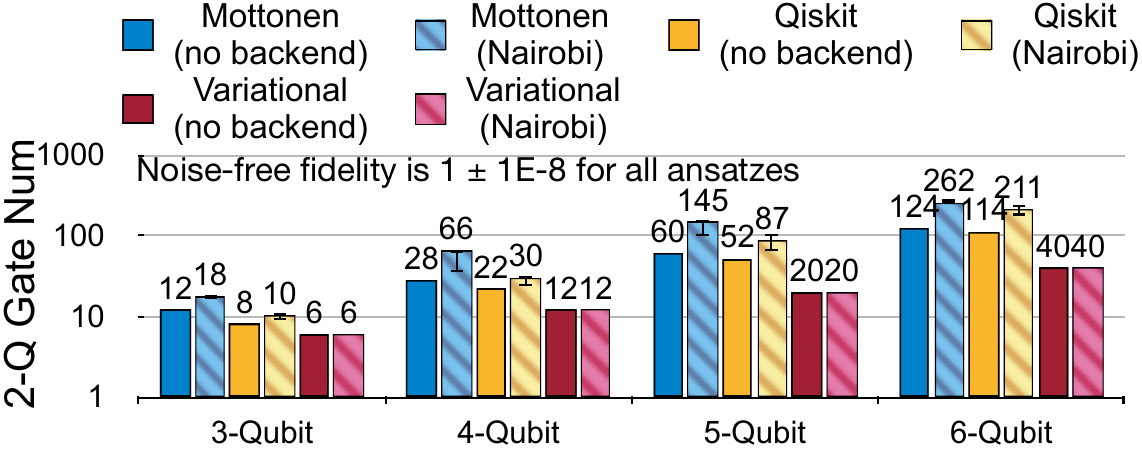}
    \caption{The VQSP ansatz has much fewer 2-Q gates but achieves similar noise-free fidelity (100\%) compared with the arithmetic decompositions (Mottonen and Qiskit).}

    \label{fig:comp_n2q}
\end{figure}

\begin{table}[t]
\centering
\renewcommand*{\arraystretch}{1}
\setlength{\tabcolsep}{14pt}
\footnotesize

\begin{tabular}{lccc}
\toprule
 & 3-Q & 4-Q & 5-Q \\
\midrule
MEAN(fidelity)& 0.9996 & 0.9988 & 0.9998 \\ 
\midrule
STD(fidelity)& 0.0013  & 0.0015 & 0.0003\\
\bottomrule

\end{tabular}

\caption{Noise-free fidelity for arbitrary states after the convergence of hardware-efficient VQSP ansatz.}

\label{tab:fid}
\vspace{-10pt}

\end{table}
Then, we compare the 2-Q gate count of VQSP circuits, achieving nearly 100\% fidelity with circuits generated by Qiskit's \texttt{initialize} method and the Mottonen method in Fig.~\ref{fig:comp_n2q}. The 2-Q gate counts for the two AD methods increase after compiling to specific hardware due to the insertion of \swap\xspace gates, while our counts remain unchanged. It is evident from the figure that variational state preparation requires fewer 2-Q gates, with reductions exceeding 6$\times$ for 6-Q states. Thus, we conclude that variational approaches are preferable for generating small-size, high-fidelity circuits.

\subsection{Noise-Aware Back-Propagation with \name}
\label{subsec:hybrid}
\begin{algorithm}[t]
    \SetAlgoLined
    \SetKwInOut{Input}{Input}
    \SetKwInOut{Output}{Output}
    \Input{Training objective $\mathcal{L}$, quantum real machine execution function $f(\cdot)$, classical simulation function $f'(\cdot)$, initial parameters $\theta^0\in\mathbb{R}^n$, initial learning rate $\eta^0$, and total steps $T$.
    }
    $\eta\gets\eta^0$\;
    \For{$t=0,1,.\cdots,T-1$}{
    \text{Execute circuit on real quantum machine}\\
    $\rho=f(\theta^t)$\;
    \text{Simulate circuit on classical simulator}\\
    $\rho'=f'(\theta^t)$\;
    \text{Objective evaluation with noisy output:}
    $\mathcal{L}(\rho)$\;
    \text{Classical backpropagation to obtain noisy gradients}\\
    $\nabla_{\theta^{t}}\mathcal{L}(\rho)=\frac{\partial\mathcal{L}(\rho)}{\partial \rho}\frac{\partial \rho'}{\partial \theta^t}$\;
    \text{Parameter update:}\\
    $\theta^{t+1}\gets\theta^t-\eta\nabla_{\theta^{t}}\mathcal{L}(\rho)$\;
     }
    \Output{Converged parameters $\theta^{T-1}$}
    \caption{\name Training with Noise-Aware Gradient Back-Propagation}
    \label{alg:hybrid_train_alg}
\end{algorithm}
    \vspace{-5pt}

\begin{figure*}[t]
    \centering
    \includegraphics[width=\textwidth]{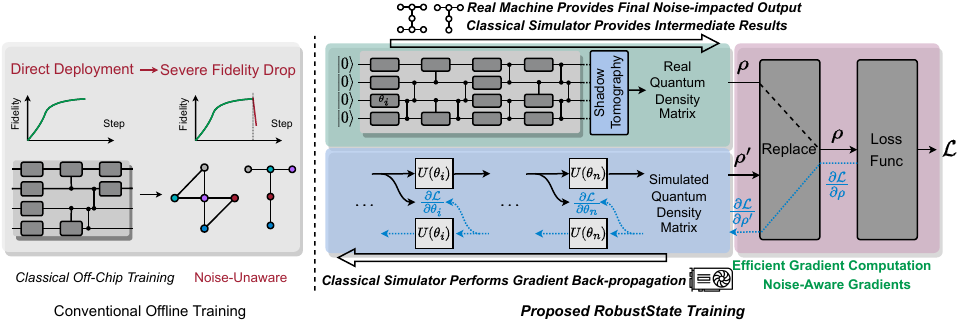}
    \caption{RobustState with real noise-aware gradients improves parameter robustness.
    }
    \label{fig:overview}
    \vspace{-5pt}
\end{figure*}

\begin{figure}[t]
    \centering
    \includegraphics[width=\columnwidth]{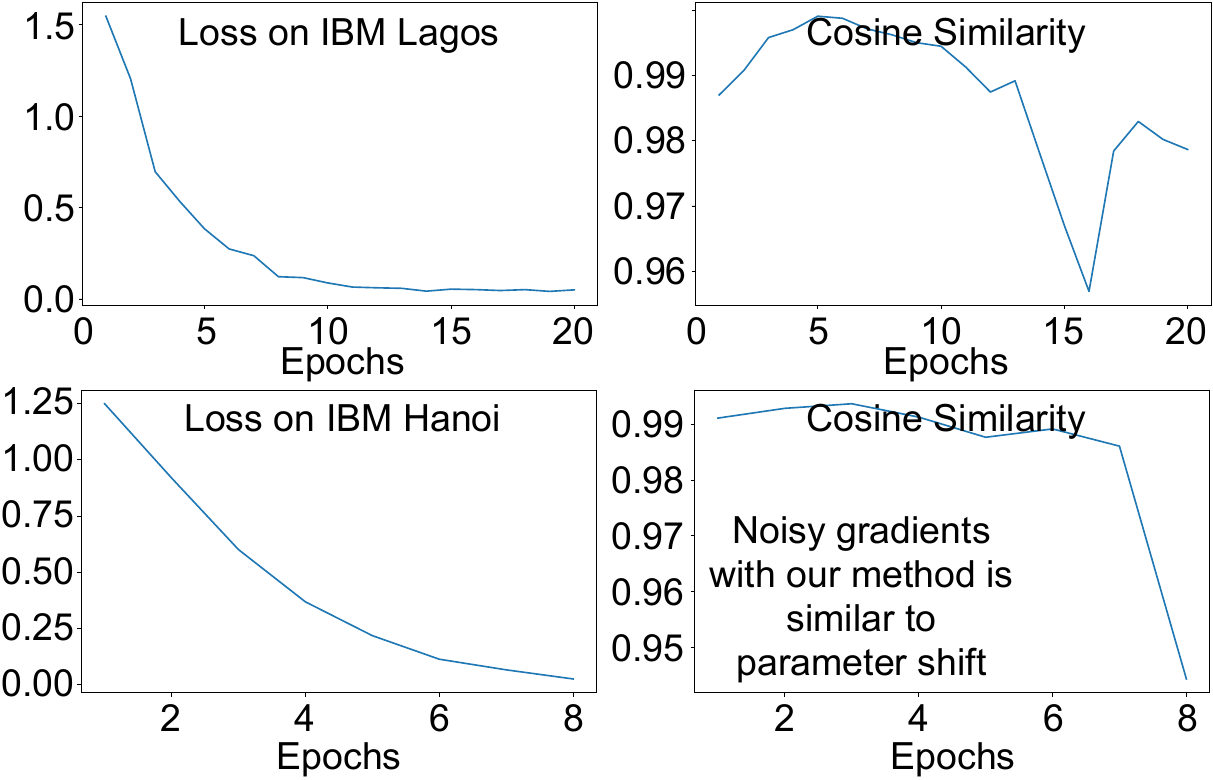}
    \caption{Noise-aware gradients approximated with \name are close to the accurate ones computed with the parameter shift rule.}
    \label{fig:gradient_similarity}
    \vspace{-5pt}
\end{figure}

Although variational circuits are smaller, they necessitate iterative parameter training. As discussed in Sec.~\ref{sec:intro}, noise-aware back-propagation is the optimal candidate for achieving high robustness and training efficiency. However, real quantum machines cannot perform back-propagation due to quantum mechanics' fundamental limits, specifically the No-Cloning Theorem~\cite{dieks1982communication, wootters1982single}, which prevents storing intermediate results necessary for back-propagation. Fortunately, the final results (density matrix) of quantum circuits can be obtained through measurements, providing abundant noise information. Inspired by \cite{wright2022physicalnn}, we employ a differentiable classical simulator to acquire intermediate results (quantum states) so that back-propagation can be performed using the noisy final output and simulated intermediate results. In Alg.\ref{alg:hybrid_train_alg}, we outline the \name training protocol, which combines quantum on-chip forward with classical simulated backward propagation for noise-aware gradient-based optimization. This process is graphically illustrated in Fig.\ref{fig:overview}. In each iteration, we first execute the quantum circuits on the real quantum device and perform tomography to obtain the density matrix $\rho$, which incorporates the effects of real noises. Next, we simulate the quantum circuit on classical computers to obtain a noiseless density matrix $\rho'$. To adjust parameters based on a specific machine's noise, we \textit{replace} the noiseless density matrix $\rho'$ with the noisy one $\rho$ to evaluate the loss function $\mathcal{L}(\rho)$.

\textbf{Noisy gradient back-propagation.} During back-prop, we adopt straight-through estimator (STE) that directly passes the noisy gradient $\frac{\partial\mathcal{L}(\rho)}{\partial \rho}$ to the theoretical noise-free path, i.e., $\frac{\partial\mathcal{L}(\rho)}{\partial \rho} \xrightarrow[]{} \frac{\partial\mathcal{L}(\rho)}{\partial \rho'}$.
Then, this estimated noisy gradient will be used to calculate derivatives for all parameters $\frac{\partial\mathcal{L}(\rho)}{\partial \theta}=\frac{\partial\mathcal{L}(\rho)}{\partial \rho'}\frac{\partial \rho'}{\partial 
\theta}$.
The fundamental reason why this gradient replacement works is that the quantum noise information can be effectively coupled in the back-propagation procedure, i.e., the noisy upstream gradient $\frac{\partial\mathcal{L}(\rho)}{\partial \rho}$, to make the training process fully aware of real quantum noises.
Note that this property requires the objective $\mathcal{L}(\cdot)$ to be a nonlinear function of the noisy $\rho$.
Otherwise, $\frac{\partial\mathcal{L}(\rho)}{\partial \rho}$ will only contain noise-free terms. This methodology synergistically leverages the noise awareness of real quantum machines and the differentiability of classical simulators for parallel, highly efficient noise-aware training. 

We will illustrate this hybrid training idea with a concrete example. In our experiments, we choose the loss function to be $\mathcal{L}=\sqrt{\tr((\rho-\hat\rho)^2)}$, then $\nabla_{\theta}\mathcal{L}=\tr(\frac{(\rho-\hat\rho)}{\mathcal{L}}\frac{\partial \rho'}{\partial \theta})$. After the parameter update, the state becomes
\begin{equation}
    \rho_{t+1}=\rho_{t}-\sum_\theta\frac{\eta}{\mathcal{L}_t}\tr((\rho_t-\hat\rho)\frac{\partial \rho'}{\partial \theta_t})\frac{\partial \rho}{\partial \theta_t}+O(\eta^2),
\end{equation}
then the loss function becomes
\begin{equation}
\begin{split}
        &\mathcal{L}_{t+1}^2-\mathcal{L}_t^2=\\-&2\sum_\theta\frac{\eta}{\mathcal{L}_t}\tr((\rho_{t}-\hat\rho)\frac{\partial \rho'}{\partial \theta_t})\tr((\rho_{t}-\hat\rho)\frac{\partial \rho}{\partial \theta_t})+O(\eta^2).
\end{split}
\end{equation}
As long as $\frac{\partial \rho}{\partial \theta_t}\approx \frac{\partial \rho'}{\partial \theta_t}$ and the learning rate $\eta$ is small enough, the two $\tr(\cdot)$ operators will have the same sign with high probability and $\mathcal{L}_{t+1}<\mathcal{L}_t$.

We further compare hybrid training approximated gradients with the real gradients estimated by parameter shift to show experimental evidence of our method's effectiveness. We build a 3-Q ansatz with 2 \rx gates on qubits 0 and 2 and an \ry gate on qubit 1, followed by three \rzx gates connecting qubits 0 and 1, 1 and 2, 2 and 0. So there are 6 trainable parameters in total. After each training step, we compare the cosine similarity of gradients between the two methods. Note that we adjust the loss function from comparing the state vector to comparing the expectation values of the Pauli-Z measurement. As in Fig.~\ref{fig:gradient_similarity}, the similarities are higher than 0.95. Therefore, our method can provide accurate noise-aware gradients while reducing the  $\mathcal{O}(\#\text{parameters})$ circuit executions in parameter shift to $\mathcal{O}(1)$ complexity per round. 

The \name is a plug-and-play approach that can be used. We note that our noise-aware gradients may not be necessary at the beginning of the training when parameters are far from convergence. Thus, in real experiments, we can combine this noise-aware training with pure classical training by first training the parameters to convergence in a simulator and then performing parameter fine-tuning using noise-aware training to reduce the cost of running on real machines. Furthermore, by replacing the loss function, our method can be used for other variational quantum algorithms to  further boost the tasks' performance. We will illustrate this using unitary synthesis and quantum state regression below.

\begin{figure}[t]
    \centering
    \includegraphics[width=0.45\textwidth]{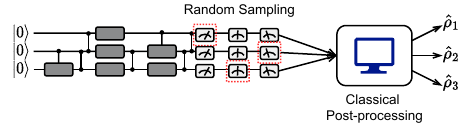}
    \caption{Sketch of the classical shadow tomography. In classical shadow tomography, we measure on random bases. After classical post-processing, each measurement result generates a snapshot of $\rho$, $\hat\rho_i$, which is an unbiased estimation of $\rho$, i.e. $\mathbb{E}(\hat\rho_i)=\rho$. With sufficient number of measurements, we can approximate $\rho$ using $\frac{1}{n}\sum_i\hat\rho_i$.}
    \label{fig:shadow_sketch}
    \vspace{-5pt}
\end{figure}

\textbf{Quantum tomography.} To get feedback from real machines, our method needs to know the prepared state $\rho$ on real machines while optimizing parameters to suit the specific noise pattern. This can be done by quantum state tomography. Among many kinds of tomography methods, we choose to use the classical shadow tomography \cite{huang2020predicting} to estimate the prepared state as shown in Fig.~\ref{fig:shadow_sketch}. Compared with traditional tomography methods like the one implemented in Qiskit \cite{smolin2012efficient}, classical shadow tomography provides an unbiased estimation of the state $\rho$ with any number of measurements. This feature allows us to sample some state bases and update the parameters without performing a full state tomography with $3^n$ circuits. This sampling is an analog of SGD and can relieve the scalability issue, as discussed in the last paragraph of Sec.~\ref{subsec:scalability}. We also adopt the readout error mitigation technique proposed in~\cite{maciejewski2020mitigation}, which prevents ansatzes from minimizing the loss function by overfitting the readout error pattern.

\subsection{Extensive Applicability of \name}
As discussed in the previous subsection, \name requires no assumption on the ansatz structure. It applies to all kinds of quantum operations with analytical formulations, while the parameter shift rule is only narrowly applicable to gates whose unitary has a structured eigenvalue~\cite{mitarai2018quantum}. So our method can be directly combined with existing ansatz design such as Xanadu~\cite{arrazola2019machine} and QuantumNAS~\cite{wang2022quantumnas} to boost their fidelity. It can also be utilized on pulse-level ansatz to further compress the circuit depth.

\begin{figure}[t]
    \centering
    \includegraphics[width=\columnwidth]{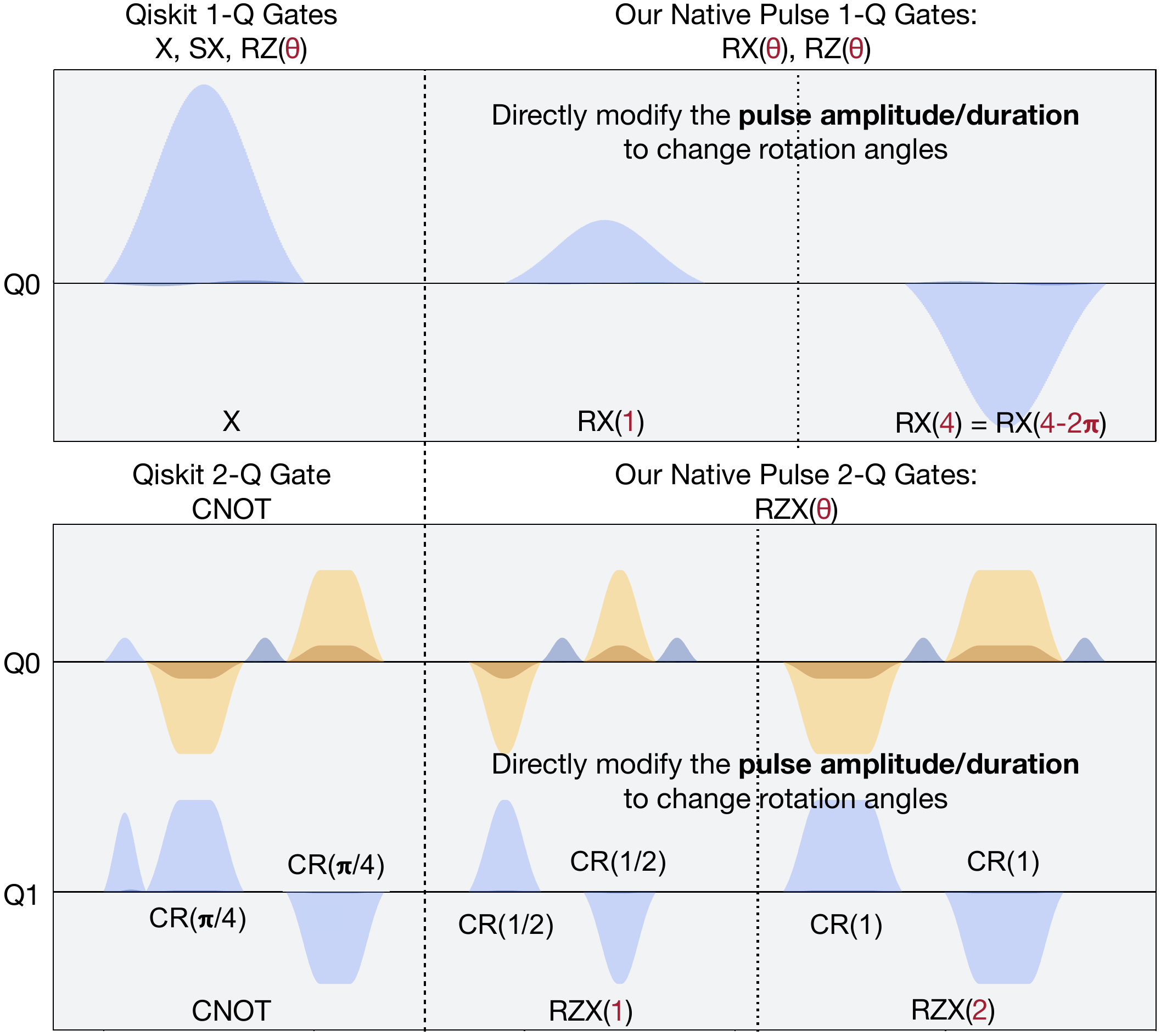}
    \caption{Pulse schedules for \xgate, \rx($1$) and \rx($4$) (top); \cnot, \rzx ($1$) and \rzx ($2$) (bottom). For IBM's quantum computers, calibration is only performed for \xgate, \sx, and \cnot, so we need to adjust the pulse shape to generate \rx($\theta$) and \rzx($\theta$) for arbitrary angles.}
    \label{fig:native_pulse_gate}
    \vspace{-5pt}
\end{figure}

Pulse-level control introduces more parameters compared to gates, which enables abundant opportunities for a more compact ansatz~\cite{gokhale2020optimized}. We can generate two more parameterized basis gates, the $\rx(\theta)$ and $\rzx(\theta)$ gates which are shown in Fig.~\ref{fig:native_pulse_gate}, without any calibration cost using pulse level control.
To implement $\rx(\theta)$ from existing calibration data, we retrieve the pulse shape of the pre-calibrated $\xgate$ gate and adjust the pulse amplitude by setting the area under the curve proportional to $\theta$. According to the basic principle of quantum dynamics, it is an approximation of $\rx(\theta)$. 
Similarly, we retrieve the pulse shape of the pre-calibrated $\crgate$, and adjust the area under it to implement $\rzx(\theta)$ for any $\theta$. 

Using a native pulse gate set has several benefits. An arbitrary single qubit rotation will be decomposed to up to $5$ native gates for IBM's default implementation (\rz,\sx,\rz,\sx,\rz). However, with native pulse gates, we can reduce this number to $3$ (\rz,\rx,\rz) \cite{9251970}. With regards to $\rzx$ gates, the entanglement operation between qubits can be precisely controlled. Thus, the circuit requires shorter runtime and has more parameters with better expressivity.

It should be noted that changing the amplitude of the pre-calibrated pulse only provides an approximation of rotation gates. Due to the imperfection of classical control and the influence of higher energy-level in superconducting quantum computers, the system is not entirely linear~\cite{gambetta2011analytic}. For example, the amplitude of $\sx$ might not be exactly half that of the $\xgate$ gate. Take IBMQ Jakarta as an example. The pulse amplitude of $\sx$ and half the amplitude of $\xgate$ has about $0.9\%$ relative difference, which introduces a detectable amount of coherent error when simply adjusting the pulse proportionally without any fine-tuning. Thankfully, our noise-aware training method in~\ref{subsec:hybrid} can automatically correct these coherent errors. With shorter pulse duration, the fidelity of the state can be improved.

\section{Evaluation}

\subsection{Evaluation Methodology}
\begin{figure*}[t]
    \centering
    \includegraphics[width=\textwidth]{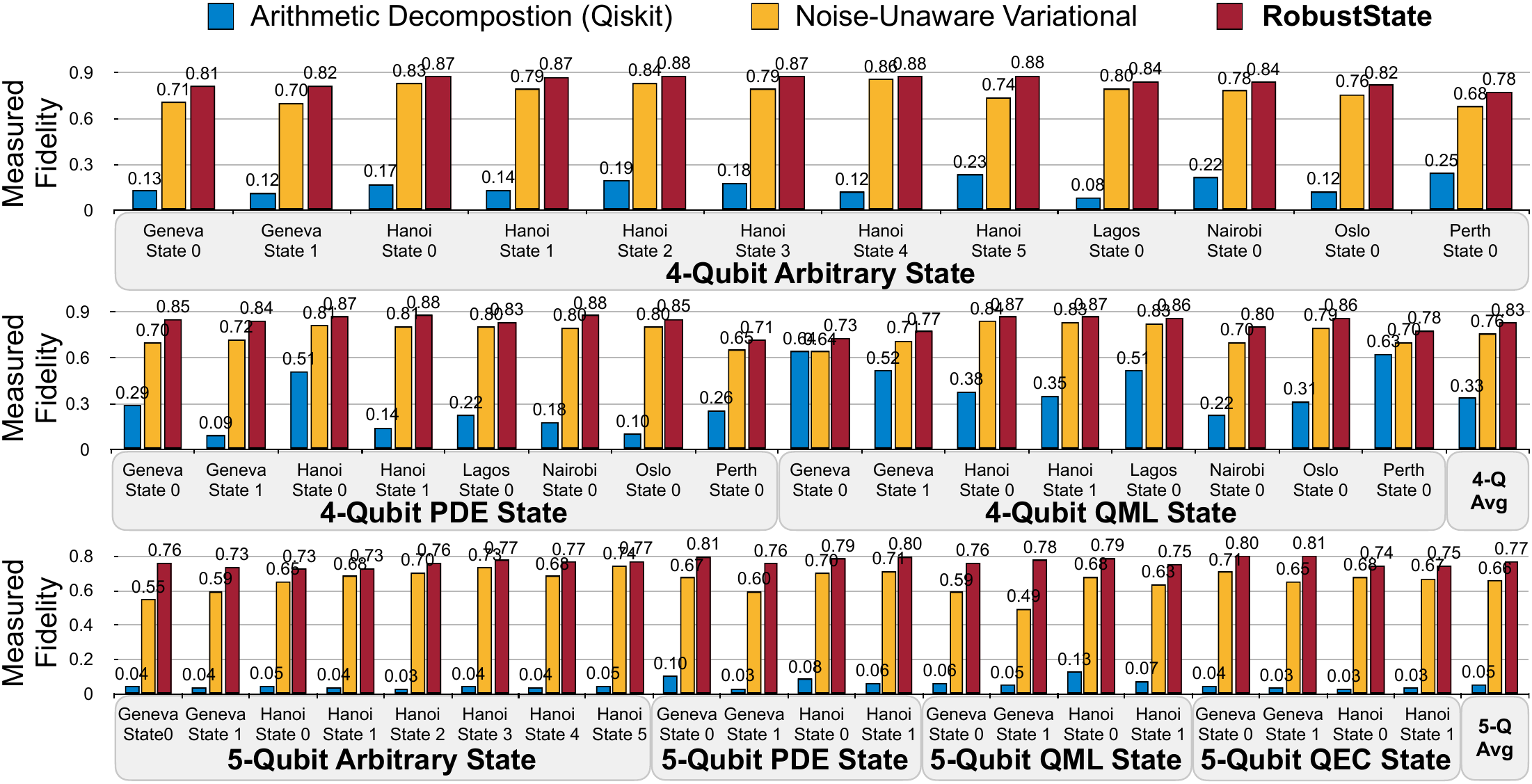}
    \caption{\name using topology-aware ansatz with \cnot\xspace gates  achieves the highest fidelity on various real machines for a number 4-Q target states when compared with the Qiskit baseline. Evaluated on real machines.}
    \label{fig:main_4q_3task}
\end{figure*}

\begin{figure*}[t]
    \centering
    \includegraphics[width=\textwidth]{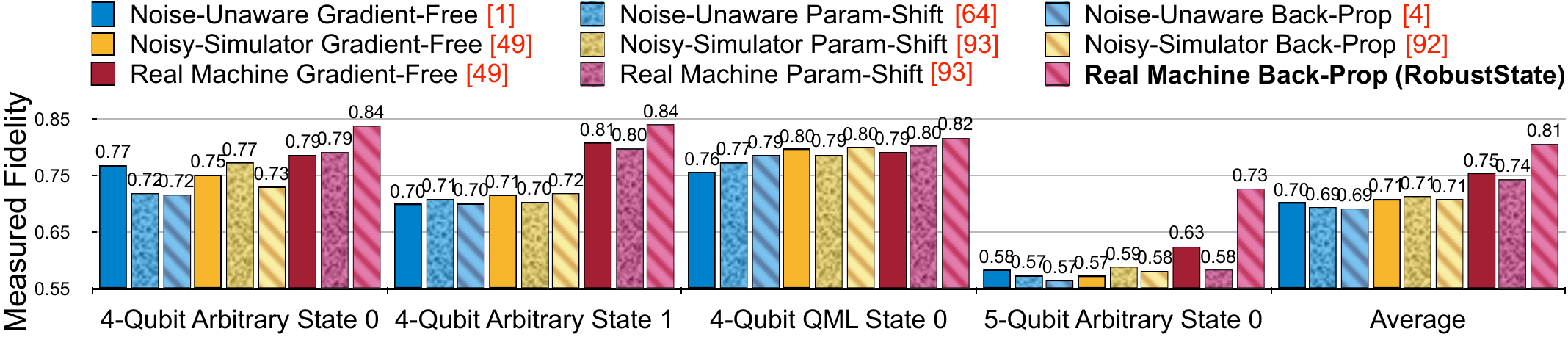}
    \caption{\name compared with eight other variational state preparation methods. \name achieved the highest fidelity for all four tasks. Evaluated on real machines.}
    \label{fig:nine_comp}
    \vspace{-5pt}
\end{figure*}

\begin{figure}[t]
    \centering
    \includegraphics[width=\columnwidth]{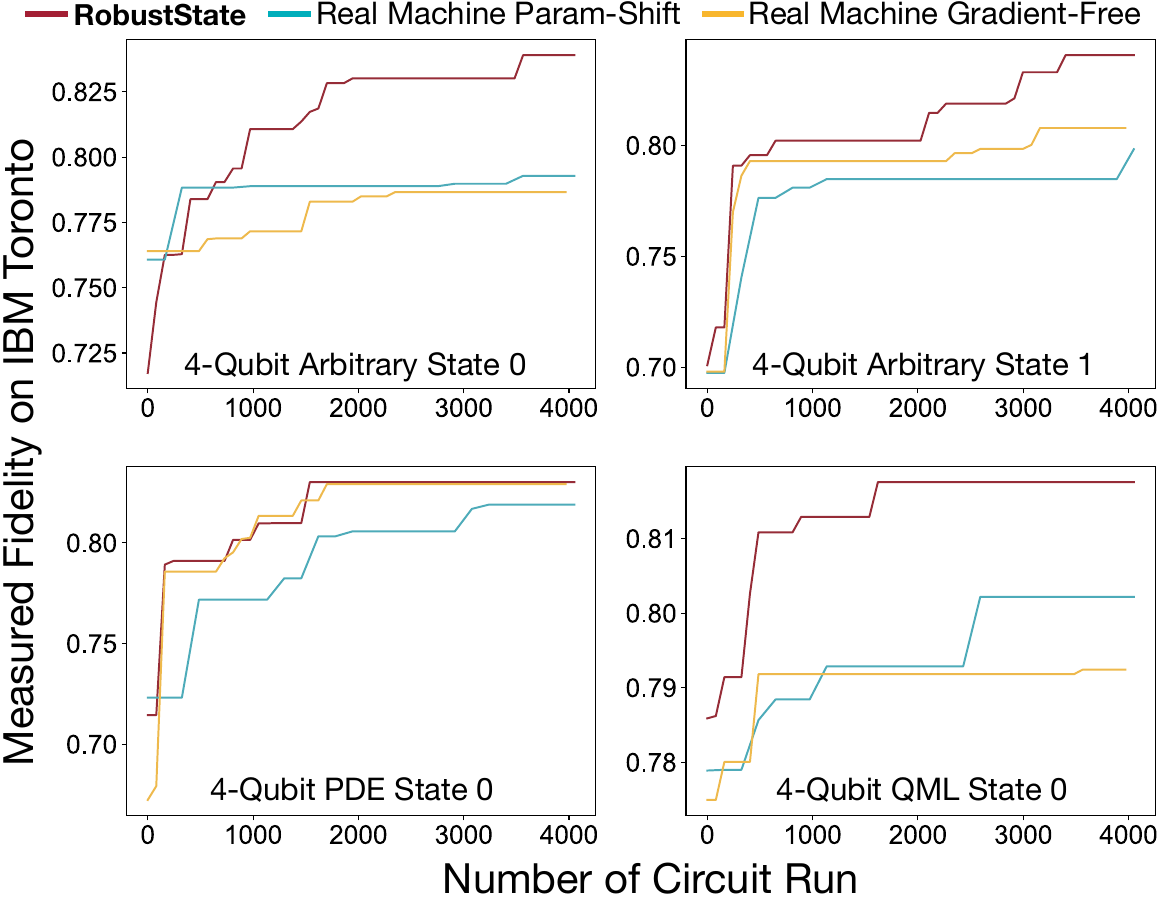}
    \caption{Comparison of three different optimization methods. Among them, \name performs the best.}
    \label{fig:curve_3methods}
\end{figure}

As explored in earlier sections, the present study examines \name in relation to existing state preparation methodologies across three distinct dimensions. Firstly, the analysis demonstrates that \name, functioning as a variational framework, surpasses the performance of AD algorithms, as illustrated in Fig.\ref{fig:main_4q_3task}. Subsequently, \name is compared to alternative variational approaches concerning training efficiency and noise-robustness, as depicted in Fig.\ref{fig:nine_comp} and Fig.~\ref{fig:curve_3methods}. The findings reveal that, by exhibiting the highest degree of training efficiency and noise-robustness, \name outperforms all other variational techniques under investigation.

\begin{figure}[t]
    \centering
    \includegraphics[width=\columnwidth]{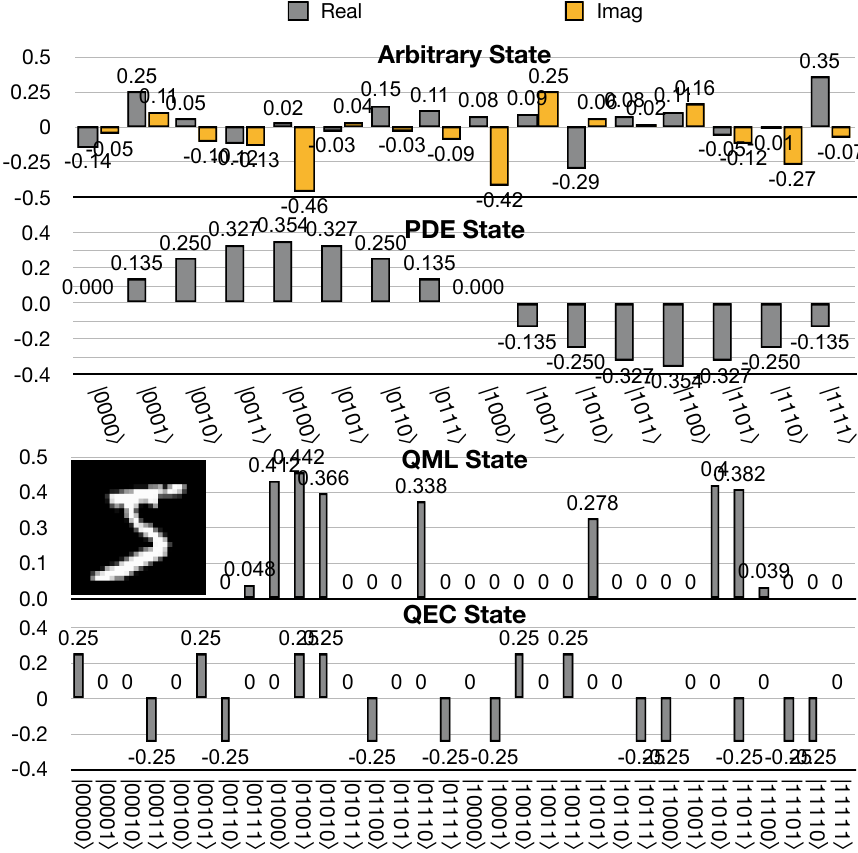}
    
    \caption{Visualization of several target states. The $y$ axis represents the amplitude on a specific basis.}
    \label{fig:target_state}
    \vspace{-5pt}
\end{figure}

Four categories of representative target states with 4 and 5 qubits are utilized, encompassing arbitrary states, partial differential equation (PDE) states, quantum machine learning (QML) states, and quantum error correction (QEC) codewords. Arbitrary states are generated following the uniform (Haar) measure. PDE states are real-valued states encoded into the amplitude of the basis states that are of interest to studying solving PDE by a variational quantum algorithm, such as the sine wave and the Gaussian distribution. The QML state encodes a classical MNIST hand-writing digit image~\cite{726791} into the state vector via amplitude encoding~\cite{nakaji2022approximate}, with the image being down-sampled, flattened, and normalized as the state. For the $5$-qubit scenario, quantum error correction (QEC) codewords for the $5$-qubit error correction code are also tested \cite{knill2001benchmarking}. Fig.\ref{fig:target_state} displays the amplitude distributions of several chosen states. Additional variational quantum algorithms, such as unitary synthesis and quantum state regression, are examined as well. Quantum state regression can be considered a model for quantum metrology\cite{marciniak2022optimal}.

\begin{figure}[t]
    \centering
    \includegraphics[width=\columnwidth]{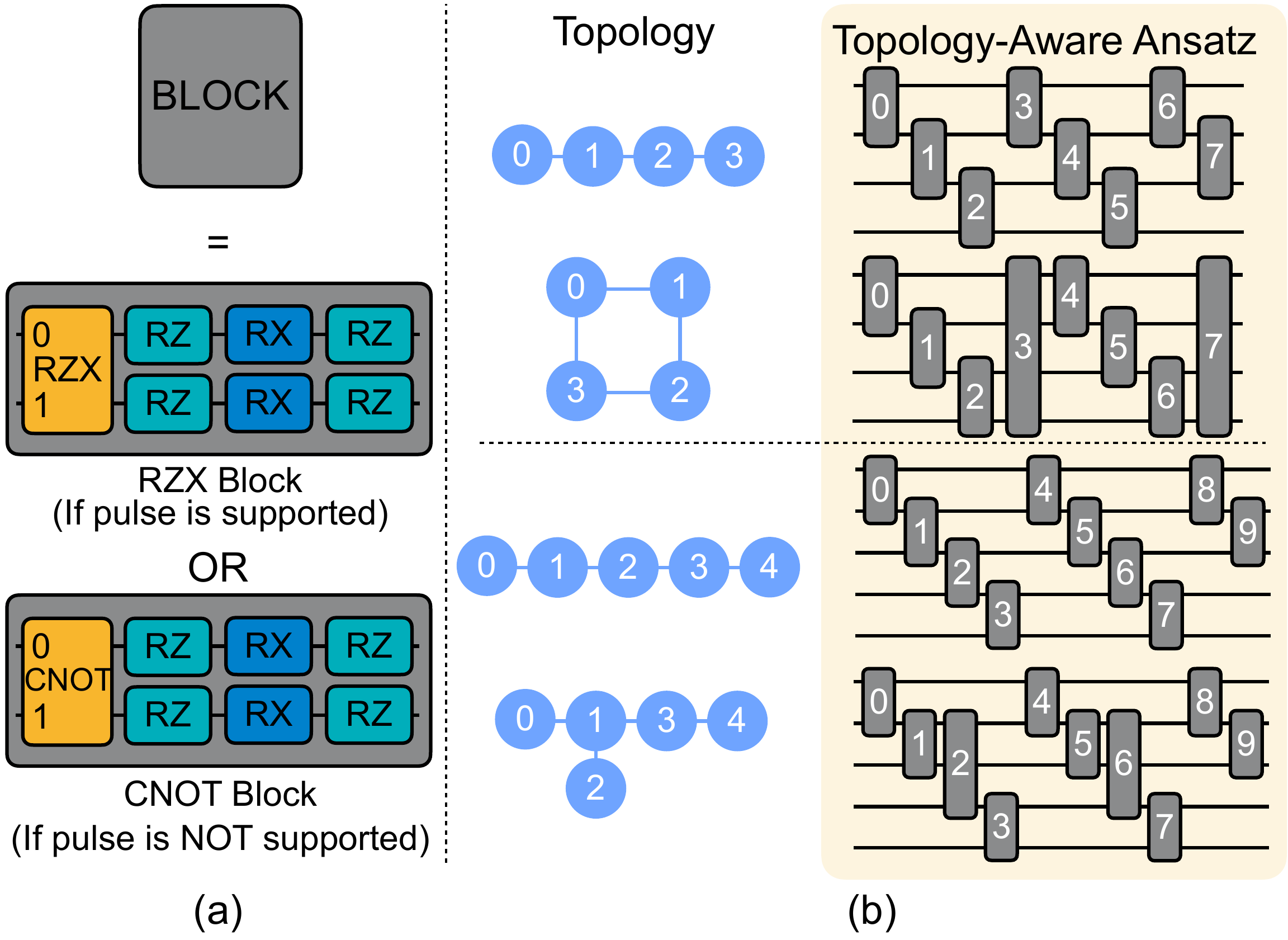}
    \caption{(a) The basic 2-Q \block \xspace used in the hardware efficient ansatz. The 2-Q gates and 1-Q gates are chosen based on the native gates of the given hardware. (b) The proposed topology-aware ansatz design with \rzx or \cnot\xspace as entangling gates. The topology-aware ansatz acts only on qubits with direct connections and attempts to execute as many \block s simultaneously as possible.}
    \label{fig:ansatz_design}
    \vspace{-5pt}
\end{figure}

\textbf{Ansatz design.} The first step of variational methods is ansatz design. Fig.~\ref{fig:ansatz_design} shows the designs of our hardware-efficient ansatz. The basic unit of the ansatz is a 2-Q \textit{block} shown in Fig.~\ref{fig:ansatz_design}(a), inspired by \cite{madden2022}. In our experiments, we use the \rzx gate as the 2-Q entangling gate on IBM machines that support pulse controls and \cnot\xspace gate on other IBM quantum machines. The \block s are only applied on neighboring qubits with direct connections to avoid any additional compilation overhead such as \swap\xspace insertions.

\textbf{Pulse gate setups.} To generate $\rzx(\theta)$ gates, we use Qiskit's \texttt{RZXCalibrationBuilder} and implement our own \texttt{RXCalibrationBuilder} accordingly for $\rx(\theta)$ gates. In addition to these two additional transpiling arguments, we also specify the virtual to physical qubit mapping, so our topology-aware ansatzes will work as expected. All the other options are set to Qiskit's default.

\textbf{Training setups.}
We use Adam optimizer with learning rate $5\times 10^{-3}$. The loss function is $\sqrt{\tr((\rho-\hat\rho)^2)}$, where $\rho$ is the target density matrix and $\hat\rho$ is state generated from the ansatz. We train the ansatz for a total of $550$ steps. For the first $500$ steps, $\hat\rho$ is obtained from a classical simulator, so the training is noise unaware; for the last $50$ steps, the training is noise aware, and $\hat\rho$ is obtained from the tomography results.

\textbf{Tomography setups.}
Unless otherwise stated, we use classical shadow tomography with all the bases measured to improve the accuracy of tomography. That is $3^4$ bases for 4-Q states and $3^5$ bases for 5-Q states. For each basis, we repeat for $1024$ shots. Due to limited shots, the estimated fidelity has a standard deviation of about $0.006$, according to our simulation results.

\subsection{Experiment Results}

~~~\textbf{Variational v.s. Arithmetic.} As a baseline, we compare \name to the \texttt{initialize} method provided by Qiskit~\cite{qiskit}, which is the only integrated arbitrary state preparation method in the Qiskit library. The \texttt{initialize} function in Qiskit is implemented based on an analog of Quantum Shannon Decomposition~\cite{shende2006}. We transpile the state preparation circuits generated by both methods with the highest \texttt{optimization\_level=3}.

Fig.~\ref{fig:main_4q_3task} shows the measured state preparation fidelity of \name on 3 kinds of states and 6 quantum machines with no pulse supports for 4 qubits and 5 qubits, respectively. Arithmetic decomposition tests the Qiskit baseline; Noise-unaware VQSP tests the classically trained VQSP, and \name is trained with real machine noise. For the tested 4-Q states, \name can achieve 83\% fidelity and a 50\% improvement on average over the Qiskit baseline.

The improvement is even more significant for 5 qubits. The Qiskit baseline averages at 5\% fidelity while \name achieved 77\% fidelity -- a 72\% improvement over the Qiskit baseline. These results demonstrate that, as we speculated in previous sections, VQSP algorithms perform much better than AD ones on NISQ. Moreover, \name can further improve the fidelity achieved by conventional VQSP.

\textbf{Additional arithmetic decomposition algorithm and error mitigation.} We perform additional baseline tests with the Mottonen algorithm~\cite{mottonen2004} implemented by PennyLane~\cite{bergholm2018pennylane} as an alternative baseline algorithm to the Qiskit baseline, as well as SWAP-based BidiREctional (SABRE) heuristic search
algorithm~\cite{li2019tackling} included by the Qiskit library used to optimize the qubit mapping to reduce \swap\xspace gate count. SABRE was implemented on both the Qiskit baseline and the Mottonen baseline. As shown in Table~\ref{tab:baselines}, \name outperforms Mottonen and Qiskit on average by 53.6\% and 42.5\% respectively. Even after applying SABRE, \name still outperforms Mottonen and Qiskit on average by 47.0\% and 35.1\%, respectively.

\begin{table}[t]
\centering
\renewcommand*{\arraystretch}{1}
\setlength{\tabcolsep}{4pt}
\footnotesize

\begin{tabular}{lcccc}
\toprule
Fidelity & Arbitrary & PDE & QML & Avg. \\
\midrule
Mottonen~\cite{mottonen2004, bergholm2018pennylane} & 0.156  & 0.175 & 0.269 &0.200 \\
Mottonen+SABRE~\cite{mottonen2004, bergholm2018pennylane, li2019tackling}   & 0.099 &  0.401 & 0.299 &0.266 \\
\midrule
Qiskit~\cite{ibm_2021} & 0.176 & 0.277 & 0.481 &0.311 \\
Qiskit + SABRE~\cite{li2019tackling} & 0.262 &  0.266 & 0.626 &0.385 \\
\midrule
\textbf{Ours}  & \textbf{0.777} & \textbf{0.713} & \textbf{0.718} & \textbf{0.736} \\

\bottomrule
\end{tabular}
\caption{\name can out-perform arithmetic decomposition methods even when noise mitigation technique such as SABRE is applied.}

\label{tab:baselines}
\vspace{-5pt}

\end{table}
\textbf{Comparison to other optimization methods.}
As introduced in Fig.~\ref{fig:teaser}, we compare \name with all eight other optimization methods. We choose the state-of-the-art optimizer, the Nelder-Mead method provided by scipy with the default setting~\cite{gao2012implementing} for gradient-free optimizations, and for parameter shift, we choose the learning rate to be $5e-3$. As shown in Fig.~\ref{fig:nine_comp}, \name achieves the highest fidelity for all four tasks tested and outperforms the second-best variational approach by 6\% on average. This result shows that, with the highest training efficiency and noise-robustness, \name is better than all other variational state preparation.

We then show the training curves of \name, parameter shift optimization and gradient-free optimization on real machines in Fig.~\ref{fig:curve_3methods}, since these are the two methods trained using real machines. For a fair comparison, we show the fidelity improvement w.r.t. the number of circuit executions. As Fig.~\ref{fig:curve_3methods} shows, our optimization method outperforms gradient-free optimization in 3 out of 4 tasks, and have equaling performance for the last task, and outperforms parameter shift in all the four tasks.

\begin{figure}[t]
    \centering
    \includegraphics[width=\columnwidth]{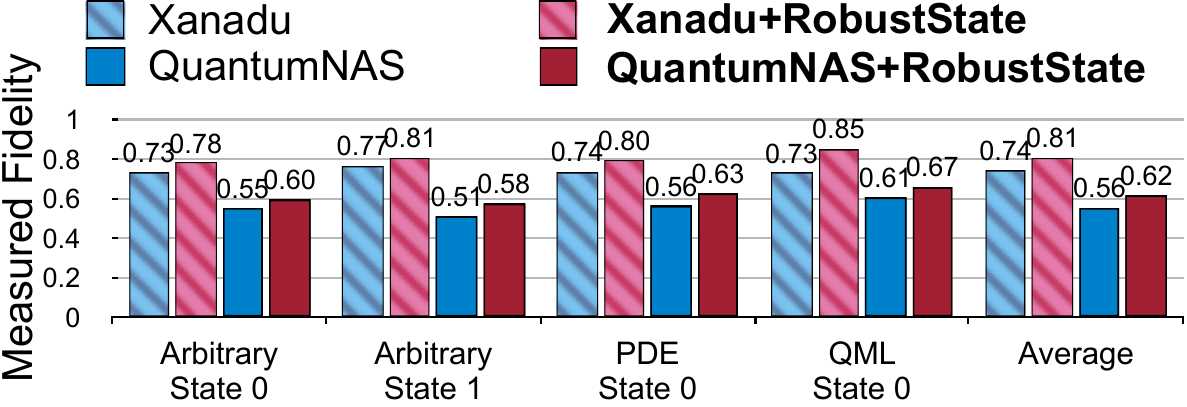}
    \caption{\name can be combined with existing ansatzes and improve their fidelity.}
    \label{fig:ansatzes}
    \vspace{-5pt}
\end{figure}

\textbf{Applicability to other ansatzes.}
As a framework for state preparation problems, we use Fig.~\ref{fig:ansatzes} to show that \name is applicable to other variational ansatzes. We apply \name on the ansatzes proposed in QuantumNAS~\cite{wang2022quantumnas} and by Xanadu~\cite{arrazola2019machine}. In the QuantumNAS ansatz, each block contains $8$ layers, the gate count of each layer is [2, 2, 4, 4, 1, 2, 2, 1], and gate type is \rx, \ry, \rz, \cnot, \rx, \ry, \rz, \cnot, respectively. We use $4$ blocks to ensure convergence. The Xanadu ansatz is originally designed for photonic quantum computers, so we adopt the design philosophy of the ansatz and implement it on superconducting quantum computers. Each block of the ansatz consists of one layer of neighboring \cnot\xspace gates as the entangling layer, one layer of single-qubit rotations, then another entangling layer, and two more layers of single-qubit rotations. We use $3$ blocks to ensure convergence. As shown in Fig.~\ref{fig:ansatzes}, on average \name improves the fidelity of the Xanadu ansatz and QuantumNAS ansatz by 7\% and 6\%, with a final fidelity of 81\% and 62\%, respectively. 
\begin{figure*}[t]
    \centering
    \includegraphics[width=\textwidth]{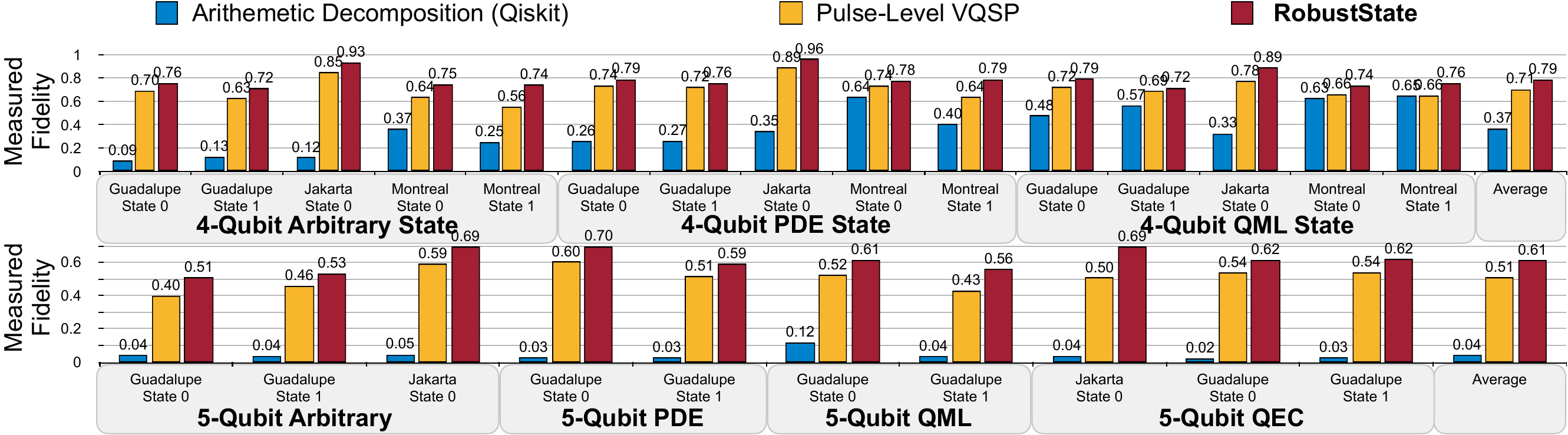}
    \caption{On three machines, IBMQ Guadalupe, IBMQ Jakarta, and IBMQ Montreal, with pulse supports, \name using the topology-aware ansatz with \rzx gates achieves the highest fidelity for various 4-Q states when compared with the Qiskit baseline. Evaluated on real machines.}
    \label{fig:main_4q_3task_pulse}
\end{figure*}

\textbf{Pulse ansatz.} Fig.~\ref{fig:main_4q_3task_pulse} shows the \name performance for 4-Q states and 5-Q states on three machines with pulse supports using \rzx blocks. We use Qiskit as our baseline here as well. \name on pulse level VQSP improves the average fidelity by 8\% and 10\% for 4-Q and 5-Q, producing final fidelities of 79\% and 61\%, respectively. Furthermore, \name with pulse outperforms the baseline by 42\% and 57\% for 4-Q and 5-Q, respectively.

\begin{figure}[t]
    \centering
    \includegraphics[width=\columnwidth]{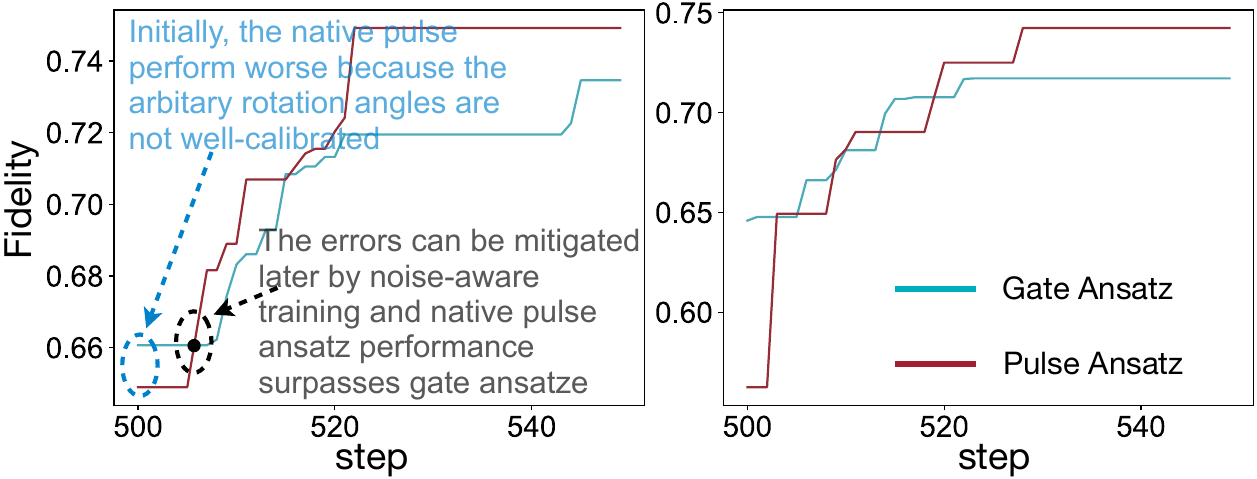}
    \caption{Training curves for the last 50 steps with the noise-aware loss for hardware-efficient ansatz using native pulse and hardware-efficient ansatz using the non-native \cnot\xspace gates. The first 500 steps are not included since they only use classical noise-free training.}
    \label{fig:pulse_vs_nopulse}

\end{figure}

\begin{table}[t]
\centering
\renewcommand*{\arraystretch}{1}
\setlength{\tabcolsep}{6pt}
\footnotesize

\begin{tabular}{lccc}
\toprule
 & Pulse & Gate&RobustState+Pulse \\
\midrule
 Low Coherent Error&  \xmark  &  \cmark  &  \cmark \\ 
 Low Incoherent Errors  & \cmark &  \xmark &  \cmark \\
\bottomrule

\end{tabular}

\caption{\name has the unique advantages of both low coherent and incoherent errors when combined with pulse level ansatz.}

\label{tab:pulse}
\vspace{-5pt}

\end{table}

An interesting phenomenon is that sometimes the noise-free trained circuit compiled to non-native basis gates performs better than that compiled to native pulse gates. This might be caused by the additional coherent error introduced by tuning pulses. Nevertheless, pulse ansatz enables \textit{better optimization space} so the final fidelity can be higher. Fig.~\ref{fig:pulse_vs_nopulse} shows the fidelity of noise-aware training with gate ansatz v.s. with pulse ansatz for two arbitrary states on the IBMQ Montreal machine. We can clearly see that the fidelity of the pulse ansatz surpasses the non-native gates in the middle of training. This shows that by combining pulse ansatz and the noise-aware gradients together, we can reduce both coherent and incoherent errors to achieve the best fidelity of quantum states, as illustrated in Table~\ref{tab:pulse}.

\begin{table}[t]
\centering
\renewcommand*{\arraystretch}{1}
\setlength{\tabcolsep}{10pt}
\footnotesize

\begin{tabular}{lcc}
\toprule
Task & Baseline & \textbf{Ours} \\
\midrule
Unitary Synthesis Jakarta & 0.845 & \textbf{0.868} \\
Unitary Synthesis Toronto & 0.858 & \textbf{0.940} \\
Unitary Synthesis Perth (1) & 0.817 & \textbf{0.834} \\
Unitary Synthesis Perth (2) & 0.798 & \textbf{0.821} \\

\midrule
Quantum State Regression (1) Loss & 0.167 & \textbf{0.147} \\
Quantum State Regression (2) Loss & 0.163 & \textbf{0.124} \\

\bottomrule
\end{tabular}
\caption{Performance of \name on other variational tasks.}

\label{tab:regression}
\vspace{-5pt}

\end{table}

\textbf{Results on other variational tasks.} 
\name is generally applicable to many other variational tasks. As an example, we test our method with quantum state regression and unitary synthesis. The quantum state regression takes an input state $\cos(\theta)\ket{000}+e^{i\phi}\sin(\theta)\ket{111}$ and predicts the values $\sin(2\theta)\cos(\phi)$ for task 1 and  $\sin(2\theta)\sin(\phi)$ for task 2. The ansatz also follows from QuantumNAS~\cite{wang2022quantumnas}. Our method can further reduce the loss for $0.03$ compared to noise-unaware training (baseline), as shown in Table~\ref{tab:regression}.
 
For unitary synthesis, we test our method for 2-Q random unitary. The ansatz we use is 6 layers of \rzx + arbitrary 1-Q gates. Compared to noise-unaware training (baseline), our method can improve the fidelity of unitaries for $3.6\%$.

\begin{figure}[t]
    \centering
    \includegraphics[width=\columnwidth]{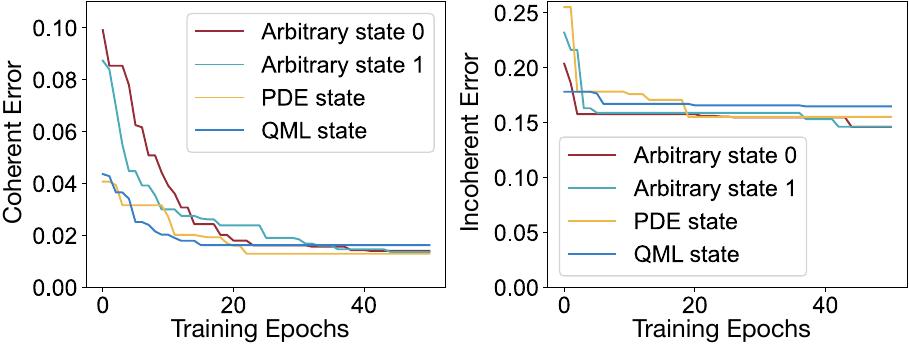}
    \caption{The coherent and incoherent errors of four state preparation tasks. Our noise-aware training method significantly reduces coherent errors.}
    \label{fig:curve_coherent}
    \vspace{-5pt}
\end{figure}

\begin{figure}[t]
    \centering
    \includegraphics[width=\columnwidth]{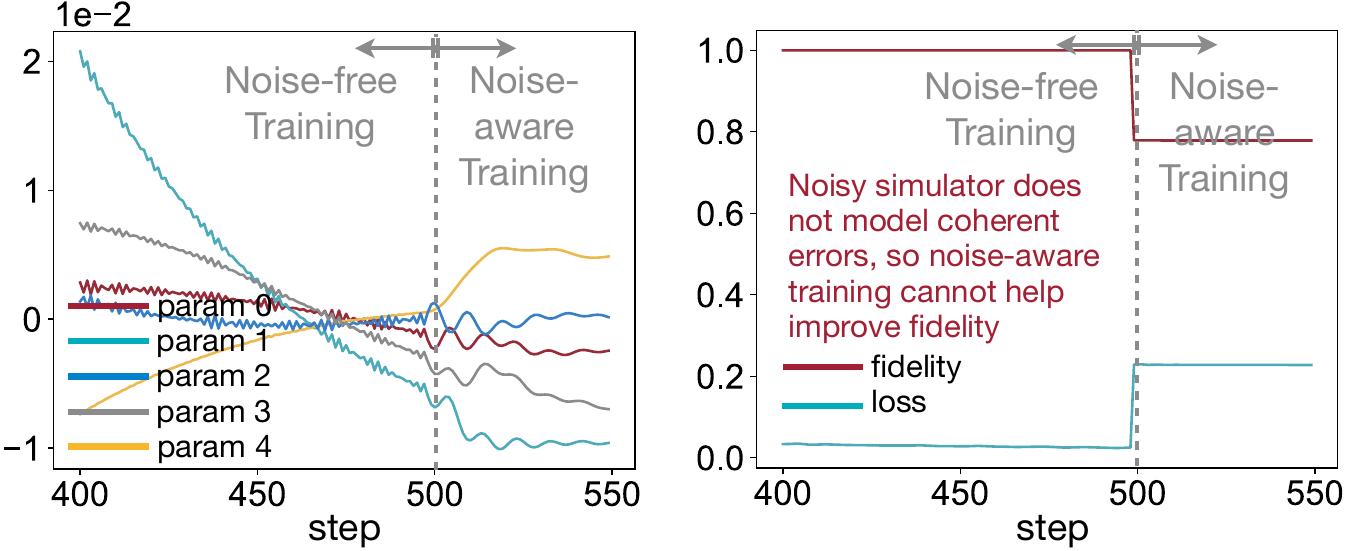}
    \caption{Training curves for selected parameters of an experiment on a noisy simulator (left); Evolution of state fidelity and loss value on the same experiment (right). The dashed lines at step 500 mark the transition from noise-free training to noise-aware training.
    }
    \label{fig:noisy_simulator}
\end{figure}

\subsection{Result Analysis}\label{subsec:results}

~~~\textbf{Where does our advantage come from?}
The noise-aware training part of our framework is targeted to eliminate coherent errors by fine-tuning the parameters. To show this, we separate the coherent error from the incoherent error in Fig.~\ref{fig:curve_coherent}. For the four states, our method can reduce the coherent errors by at least $62\%$ and up to $86\%$. However, for incoherent error, it is only reduced by as low as $7\%$ and no more than $39\%$.

This can be further illustrated with a simulator experiment.
We conduct our method with similar settings but on a noisy simulator instead of real quantum computers. As shown in Fig.~\ref{fig:noisy_simulator}, the parameters do not change substantially when training with noise-aware gradients, and the fidelity is only marginally improved. This is not surprising, as the Qiskit noisy simulator only models incoherent errors (This is not to say that the simulator cannot simulate coherent error, but that the current noise model used by the noisy simulator has insufficient information about coherent errors. During calibration, if we find an over-rotation error, we will immediately fix the error by reducing the pulse amplitude instead of writing that information into the noise model). Noise-aware gradient, by itself, does not change the circuit depth, so it cannot optimize the parameters for a simulator without any coherent error.

\begin{figure}[t!]
    \centering
    \includegraphics[width=\columnwidth]{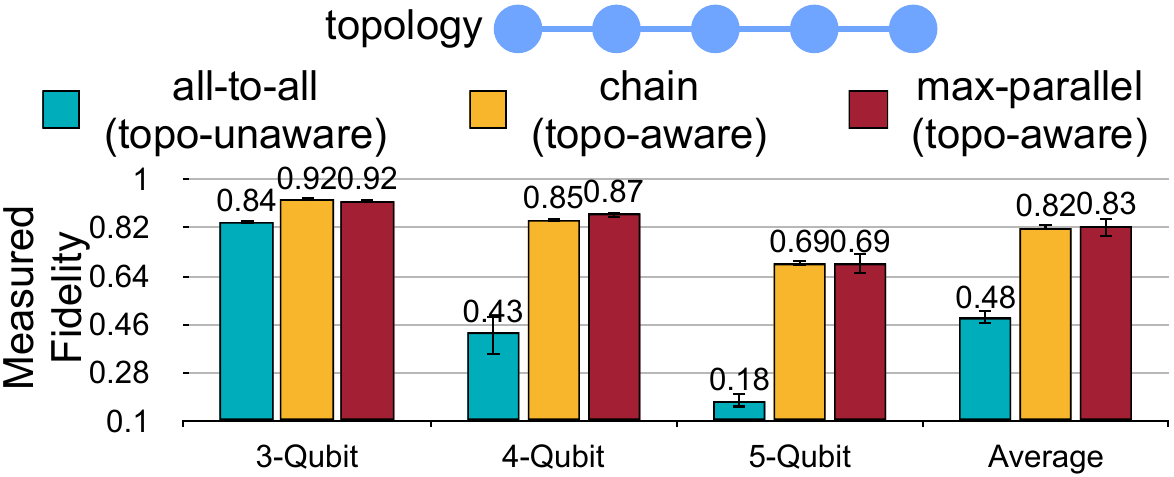}
    \caption{The topology-aware ansatz compared with the topology-unaware ansatz with classically optimized parameters run on a linear array of qubits of IBM Hanoi. 
    }
    \label{fig:compare_topo}
\end{figure}

\begin{table}[t]
\centering
\renewcommand*{\arraystretch}{1}
\setlength{\tabcolsep}{12pt}
\footnotesize

\begin{tabular}{lcc}
\toprule
Topology & Noise-Unaware Back-Prop & RobustState \\
\midrule
I&0.828&\textbf{0.875}\\ 
\midrule
T&0.674&\textbf{0.797}\\
\bottomrule

\end{tabular}

\caption{Fidelity for Different Topologies}

\label{tab:topologyandfidelity}

\end{table}

\textbf{Whether the topology-aware ansatzes perform better?}
In the noise-free simulation of different ansatzes, both topology-aware and topology-unaware ansatzes converge to the target state with a similar number of \block s. We show the results of running these state preparation circuits on a set of linearly connected qubits from the IBM Hanoi machine in Fig.~\ref{fig:compare_topo} using 3, 4, and 5 qubits. The fidelities dropped from 3 to 5 qubits as the deeper circuits would introduce more noise. For the topology-aware ansatz, fidelity went from $92\%$ to $69\%$, while for the topology-unaware ansatz, the fidelity went from $84\%$ to $18\%$, showing importance of topology awareness.

\textbf{Whether our method generalizes to different topologies?} We test our method with different topology (Table~\ref{tab:topologyandfidelity}), the $I$ type (the first row of Fig.~\ref{fig:ansatz_design}) and $T$ type (the last row of Fig.~\ref{fig:ansatz_design}). Both of them use the topology-aware ansatz. It can be seen from the table that our method works well for different topologies. And for the topology-aware ansatz, the $I$ topology works better since it has a shallower circuit.

\begin{figure}[t]
    \centering
    \includegraphics[width=\columnwidth]{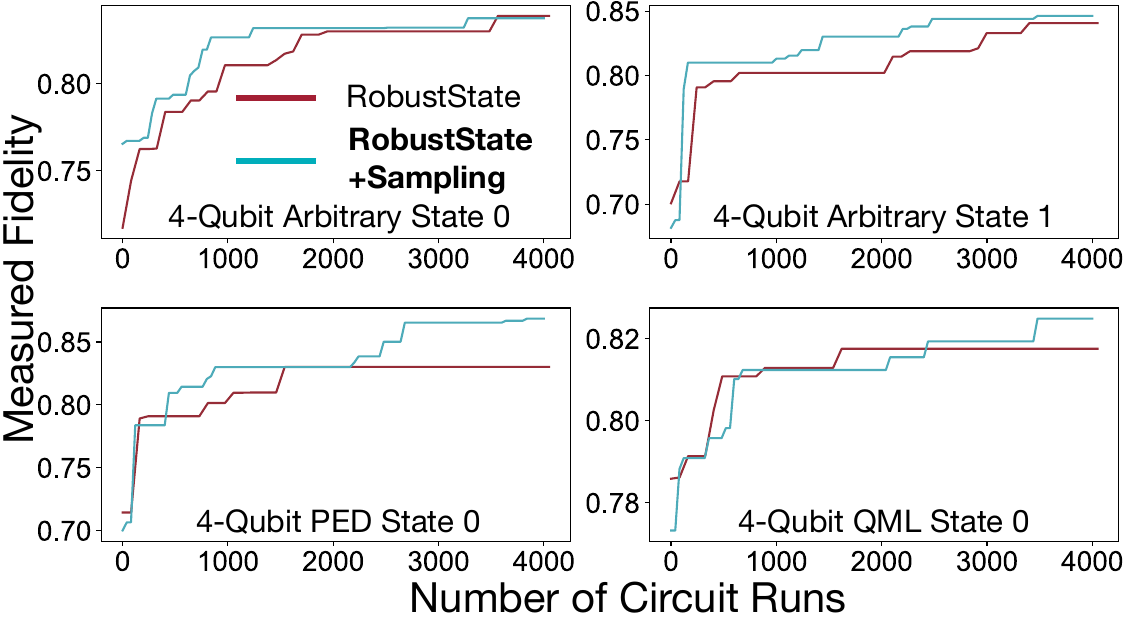}
    \caption{Comparison between full state tomography and basis sampling. Basis sampling works better than full tomography in general.}
    \label{fig:curve_sampling}
    \vspace{-5pt}
\end{figure}

\textbf{Is our method scalable?}
\label{subsec:scalability}
Preparing small-to-medium states with high fidelity is essential to many quantum algorithms in the NISQ era. For problems with small sizes, the cost of the noise-free training part is negligible. On a Mac with M1-pro processor and the TorchQuantum library~\cite{wang2022quantumnas}, training for $500$ steps takes about $6.02$s for a 4-Q ansatz with 12 $\cnot$ \textit{block}s, and $11.12$s for a 5-Q ansatz with $20$ $\cnot$ \textit{block}s. TorchQuantum supports training for up to 30 qubits, so classical simulation is not a problem for medium-sized tasks. With regards to on-chip training, the classical shadow tomography we used enables us to update the parameter in a stochastic way which can mitigate the scaling issue. We sample $40$ bases out of $81$ bases to estimate the loss function, and our experiments show that basis sampling enables faster convergence and higher fidelity. The results are shown in Fig.~\ref{fig:curve_sampling}. Though our method cannot scale up with even more qubits, preparing quantum states with a few qubits with high fidelity is already very useful for near-term NISQ and long-term fault-tolerant quantum computing.

\section{Related Work}

\subsection{Pulse-Level Quantum Computing}
Recently, quantum algorithm optimization on the pulse signal level has attracted increasing research interest~\cite{gokhale2020optimized, shi2019optimized, liang2022variational, meitei2021gate, johansson2012qutip, cheng2020accqoc, de2022pulse}. A series of works explores Quantum Optimal Control~\cite{shi2019optimized, cheng2020accqoc, de2022pulse}, which iteratively optimizes the pulse shapes of channels of the systems by minimizing the distance between the computed Unitary matrix of the pulses and the target. Another research focus is on variational pulse learning~\cite{liang2022variational, johansson2012qutip, meitei2021gate, liang2022pan} in which pulse shape parameters (such as amplitude, frequency, and duration) are directly optimized instead of the angles of rotation gates. 
\cite{gokhale2020optimized} proposes to compile the gates to the native gate sets to reduce the circuit depth. Our method improves upon theirs because our hybrid training can mitigate the unwanted coherent errors introduced by directly changing pulse shapes according to gate parameters.

\subsection{Noise-Aware Quantum Compilation}
Noise constitutes a significant challenge in NISQ machines, prompting the development of various noise-adaptive quantum compilation techniques to mitigate its effects~\cite{ravi2022vaqem, 10.1145/3503222.3507703, das2021jigsaw, 10.1145/3445814.3446743, wang2022quest, wang2023dgr, wang2023transformerqec, wang2023qpilot}. These approaches aim to reduce noise impact by addressing diverse gate errors that can be suppressed through different strategies.
Qubit mapping techniques~\cite{murali2019noise, tannu2019not, li2019tackling, molavi2022qubit, liuqucloud, zhang2021time} help minimize noise by optimizing the assignment of logical qubits to physical qubits on quantum devices. Composite pulses~\cite{merrill2014progress, low2014optimal, brown2004arbitrarily, xie2022suppressing} involve designing sequences of pulses that achieve a desired operation while mitigating errors caused by noise.
Dynamical decoupling~\cite{hahn1950spin, viola1999dynamical, biercuk2009optimized, lidar2014review, das2021adapt} techniques are employed to isolate quantum systems from their environment, reducing the impact of noise on their evolution. Randomized compiling~\cite{wallman2016noise} introduces randomness into quantum circuits to average out systematic errors, while hidden inverses~\cite{zhang2021hidden} utilize a set of quantum operations that are robust against specific noise channels.
Instruction scheduling~\cite{murali2020software, wu2020tilt} optimizes the order of quantum operations to minimize noise, and frequency tuning~\cite{versluis2017scalable, helmer2009cavity, ding2020systematic} adjusts the frequencies of qubits to reduce crosstalk and enhance gate fidelity. Parallel execution on multiple machines~\cite{stein2022eqc} leverages the resources of several quantum devices to overcome noise limitations.
Algorithm-aware design and compilation techniques~\cite{li2022paulihedral, lao20222qan, cheng2022topgen, zheng2022sncqa} tailor the compilation process to exploit specific algorithmic properties, leading to improved noise resilience. Finally, qubit-specific basis gate approaches~\cite{lin2022let, li2021software} design custom basis gates for individual qubits, optimizing their performance in the presence of noise.

\section{Conclusion}
We introduce \name, a noise-aware training framework for robust quantum state preparation. Our framework's key feature is its ability to combine noise-impacted outputs from real machines and intermediate results from simulators to perform noise-aware back-propagation. This approach results in highly noise-robust parameters and increased training efficiency. We thoroughly evaluate the \name framework on 10 real quantum machines, showcasing a reduction of over 7.1$\times$ in coherent error and achieving, on average, 50\% and 72\% fidelity improvements over baselines. We also validate our noise-aware training on other variational algorithms including unitary synthesis and state regression. Moving forward, we anticipate that the proposed noise-aware training framework can be generalized to other variational algorithms to mitigate real machine noise and serve as a valuable subroutine for numerous quantum algorithms.


\bibliographystyle{IEEEtranS}
\bibliography{main, mypaper}

\end{document}